\documentstyle[12pt,aasms4]{article}

\begin{document}

\title{Correlation between the $\gamma$-ray and the radio emissions}

\author{J.H. Fan and G. Adam}
\affil{  CRAL Observatoire de Lyon, 9, Avenue Charles André, 69 563 St-Genis-Laval Cedex, France}

\author{G.Z. Xie}
\affil{ Yunnan Observatory, Chinese Academy of Sciences, Kunming 
650011, China} 

\author{S.L. Cao}
\affil{ Department of Astronomy, Beijing Normal 
University, Beijing, China}

\author{R.G. Lin}
\affil {Center for Astrophysics, Guangzhou Normal University, Guangzhou 510400, China}

\author{Y. Copin}
\affil{ CRAL Observatoire de Lyon, 9, Avenue Charles André,
 69 563 Saint-Genis-Laval Cedex, France}

\begin{abstract}
In this paper, the correlation between the $\gamma$-ray and the radio bands is investigated.  
The results show that there is a closer correlation between the $\gamma$-ray emission and 
the high frequency ( 1.3mm, 230GHz) radio emission for maximum data than between the 
 $\gamma$-ray and the lower frequency (5GHz) radio emissions, which means that the 
 $\gamma$-ray is associated with the radio emission  from  the jet.
 
\end{abstract}

\section{Introduction}

 The most important result of the CGRO/EGRET instrument in the field of extragalactic astronomy 
 is the discovery that blazars (i.e., flat-spectrum radio quasars--(FSRQs) and BL Lac objects) emit
  most of their bolometric luminosity in the high  $\gamma$-ray ($ E > $100 MeV ) energy range.  
 Many of the $\gamma$-ray emitters are also superluminal radio sources (von Montigny et al. 1995).  
 The common properties of these EGRET-detected AGNs are the following:  The $\gamma$-ray flux is 
 dominant over the flux in lower energy bands; The $\gamma$-ray luminosity above 100 MeV ranges 
 from less than $3 \times 10^{44}$ erg/s to more than $10^{49}$ erg/s; Many of the sources are 
 strongly variable in the $\gamma$-ray band on timescales from days to months, but large flux 
 variability on short timescales of $< 1$ day has also been detected (see 0716+714 for instance, 
 Cappi et al. 1994) and the photon spectrum in the EGRET energy range (30 MeV to
 30 GeV) are generally well represented by power laws with an average  photon spectral index of 2.0.

 Various  models for $\gamma$-ray emission have been proposed: 
 (1) the inverse Compton process on the external photons ($ ECS$),  in which the soft photons 
 are directly from a nearby accretion disk ( Dermer et al. 1992; Coppi et al. 1993 ) 
 or  from disk radiation reprocessed  in some region of AGNs ( e.g. broad emission line region) 
 (Sikora et al.  1994; Blandford \& Levinson 1995); 
 (2) the synchrotron self-Compton model ($SSC$), in which the soft photons originate as 
 synchrotron emission in the jet (Maraschi et al. 1992; Bloom \& Maraschi 1992, 1993;
  Zdziarski \& Krolik 1993); 
 (3) synchrotron emission from ultrarelativistic electrons and positrons produced in a 
 proton-induced cascade ($PIC$) (Mannheim \& Biermann 1992; Mannheim 1993; Cheng \& Ding 1994).   
 From these models it is clear that the $\gamma$ -ray emission is from the jet.  Observations 
 suggest that most of the objects in the  EGRET sample show superluminal motion, which yields 
 also strong evidence that the $\gamma$-ray radiation from these objects comes from the 
 relativistic jets and is strongly beamed.

 As the models indicate, there is no consensus yet on the dominant emission process (see $3C273$ 
 for instance, von Montigny et al. 1997).  It is well known that the emission might imply various
  relations among wave bands that can be used to distinguish among a variety of
 emission mechanisms.  Dondi \& Ghisellini (1995) have studied the correlation between emission 
 in the $\gamma$-ray and in the lower energy bands, and found that the $\gamma$-ray luminosity 
 is more correlated with the radio luminosity than with other bands luminosities
  (e.g. optical and X-ray band); but M$\ddot{u}$cke et al. (1997) reported that there is no 
 correlation between the $\gamma$-ray and the radio bands.  Xie et al. (1997) found that the 
 luminosity correlation between the $\gamma$-ray and the infrared band is  closer than that
 between the $\gamma$-ray and the optical or the X-ray band.  Fan (1997) has investigated the 
 correlation between the $\gamma$-ray band and the lower energy bands by means of the multiple 
 regression method. He  found that there is an indication of a correlation between the 
 $\gamma$-ray flux and the radio flux while there is no correlation between the $\gamma$-ray 
 flux and the optical flux or between the $\gamma$-ray flux and the X-ray
 flux, and proposed that the $\gamma$-ray emission is from the $SSC$ process and that the 
 correlation between the $\gamma$-ray and the radio bands is probably due to the fact that 
 both the $\gamma$-ray and the radio emissions are beamed.  Observations show that there is 
 a  correlation between the $\gamma$-ray and radio bands (Valtaoja \& Ter$\ddot{a}$sranta 1995) 
 although there is no simple one-to-one relation between them (Pohl et al. 1996; 
 M$\ddot{u}$cke et al. 1996a).  We think that the reason for these different results comes 
 from the following factors: 1) Luminosity-luminosity correlation can not be considered as a true 
 correlation because of the known fact that luminosity depends on redshift; 2) The lower 
 frequency radio emission is not only from the jets and is variable; 3) The $\gamma$-ray 
 emissions show large flux variation (von Montigny et al. 1995, see also Hartman 1996). 
 These facts suggest that the correlation between the $\gamma$-ray and the radio bands 
 is difficult to conclude.  So, we will propose that it is necessary to use the high frequency 
 radio data to investigate the association between the $\gamma$-ray and the radio band 
 emissions.  Here we will use the observed maximum data in the $\gamma$-ray and radio bands,
 the sources are listed in table 1. In section 2, we  give the data and the correlation 
 between the $\gamma$ -ray and the radio bands; In section 3, we give some discussion.

\section{Correlation}
\subsection{Data}
 Blazars are known to be strongly variable in the $\gamma$-ray as well as in the radio 
 band on time scales of days to months (von  Montigny et al. 1995). Therefore, simultaneous
 observations should be adequate  for a correlation analysis  (M$\ddot{u}$cke et al. 1997).  
 Unfortunately, there is scarity of such simultaneous observations.  So, we can only choose
 the observed maximum high  frequency  data in the radio band at 230GHz and the observed 
 maximum data in the $\gamma$-ray band to investigate the  correlation between the 
 $\gamma$-ray and the radio emission. Radio data obtained after 1990 have been chosen 
 because this  corresponds to the operation period of EGRET.  

 In this paper, we discuss 44 $\gamma$-ray loud AGNs with available 
 high frequency radio (230 GHz) flux densities ( see Table 1). 35 are FSRQs (
 19 highly polarized quasars -- HPQs with $ P>3 \% $;  11 are lowerly polarized 
 quasars--LPQs with $ P<3\%$; and 5 objects have no available polarization measurements); 
 9 of which are BL Lac objects and are  marked with a $\dagger$.  Col.1
 gives the name of the source; Col. 2, the redshift, Col.3,  the observed maximum 
 $\gamma$ -ray photon in $10^{-7}$photon/cm$^{2}$/s with the error, Col. 4,  the spectral 
 index; Col.5,  reference for Col. 3 \& 4; Col. 6,  the radio flux in Jy at 5GHz;
 Col. 7, reference for Col. 6 (see also Comastri et al. 1997; M$\ddot{u}$cke et al. 1997); 
 Col. 8  the observed maximum  high frequency radio flux in Jy and the error, Col. 9
 references for Col. 8.  As in the paper of Comastri et al. (1997),
 the adopted $\gamma$-ray data of 1622-297 is not the peak value of 
 (210$\pm$70)$\times 10^{-7}$photon/cm$^{2}$/s (Mattox \& Wagner 1996) 
 but the data compiled by Mukherjee et al. (1997). It is found that the $\gamma$-ray spectrum
 tends to harden with increasing $\gamma$-ray flux for EGRET sources (M$\ddot{u}$cke et al. 1996b). 
 A strong correlation has also  been found for the spectral index and the integral flux above 
 100Mev for 3C273 (von Montigny et al. 1997). So, we chose the flat
 spectral index if there are more than one spectral index available for the sources 
 considered in the paper.

\begin{table*}
\caption{ A Sample of $\gamma$-Ray Loud AGNs with Available High Frequency Radio Data at 230~GHz.}
\begin{tabular}{lcccccccc} 
\hline\noalign{\smallskip}
 $Name$ & $Redshift$ & $N_{(\>100MeV)}(\sigma)$ & $\alpha_\gamma$ &  References & $f_{5GHz}$ & Ref & $f_{230GHz}(\sigma)$ & Reference \\
\noalign{\smallskip} \hline
0202+149  & 1.202  &   2.6(0.60)  & 1.50  &  F94      &  2.49 & K81  & 0.85(0.07) &T96 \\
0208-512  & 1.003  &  13.19(2.47) & 0.70  &  T95,M97  &  3.31 & K81  & 2.60(0.21) &T96 \\
0234+285  & 1.213  &   2.91(1.13) & 1.70  &  T95      &  2.36 & P82  & 2.66(0.30) &S92 \\
0235+164$^\dagger$  & 0.940  &   8.25(0.91) & 0.90  &  T95      &  2.85 & S91  & 3.82(0.31) &T96 \\
0336-019  & 0.852  &  18.62(0.76) &       &  M97      &  2.84 & K81  & 1.35(0.12) &T96 \\
0420-014  & 0.915  &   5.12(1.05) & 0.90  &  T95,M97  &  3.72 & P82  & 5.34(0.38) &T96 \\
0440-003  & 0.844  &   8.44(1.20) &       &  M97      &  3.17 & W85  & 0.78(0.07) &S92 \\
0446+112  &         &   11.3(2.06) & 0.80  &  T95      &  1.22 & K81  & 1.39(0.12) &T96 \\
0454-234  & 1.009  &   1.40       &       &  vM95     &  2.20 & L85  & 0.88(0.06) &T96 \\
0454-463  & 0.858  &   2.90       & 0.90  &  vM95     &  2.97 & K81  & 0.51(0.04) &T96 \\
0458-020  & 2.286  &   3.08(0.95) &       &  M97      &  2.04 & L85  & 0.92(0.10) &T96 \\
0506-612  & 1.093  &   0.60       &       &  vM95     &  1.50 & K81  & 0.45(0.04) &T96 \\
0521-365  & 0.055  &   3.75(1.12) & 1.2   &  T95,M97  &  9.70 & K81  & 3.98(0.32) &T96 \\
0528+134  & 2.070  &  30.76(3.46) & 1.30  &  T95,M97  &  4.30 & P82  & 4.21(0.35) &T96 \\
0537-441$^\dagger$  & 0.894  &   8.98(1.45) & 1.00  &  T95,M97  &  4.00 & S91  & 5.73(0.46) &T96 \\
0716+714$^\dagger$  &        &   4.40(1.1 ) & 0.90  &  T95,M97  &  1.12 & K81  & 3.03(0.31) &S92 \\
0735+178$^\dagger$  & 0.424  &   4.09(2.13) &       &  M97      &  3.65 & G94  & 0.92(0.11) &T96 \\
0827+243  & 0.939  &   6.81(1.44) & 1.30  & M97,F94 &  0.67 & B91  & 1.33(0.11) &T96 \\ 
0836+710  & 2.172  &   4.53(1.13) & 1.40  &  T95      &  2.67 & P82  & 0.93(0.09) &S93 \\
0851+202$^\dagger$  & 0.306  &   2.90       &       &  Sh96     &  2.70 & K81  & 2.50(0.26) &T96 \\
0906+430  & 0.670  &   3.20       &       &  C97      &  1.80 & K81  & 0.40(0.04) &S88 \\
0954+658  & 0.368  &   1.43(0.40) & 0.90  &  T95      &  1.46 & K81  & 0.54(0.05) &S88 \\
1127-145  & 1.187  &   9.27(2.29) & 1.15  &  S96      &  7.46 & K81  & 1.22(0.09) &T96 \\
1156+295  & 0.729  &  22.86(5.48) & 1.0   &  T95      &  1.65 & G94  & 0.83(0.08) &T96 \\
1219+285$^\dagger$  & 0.102  &   1.7        & 0.40  &  vM95     &  0.97 & G91  & 0.19(0.04) &T96 \\
1222+216  & 0.435  &   8.29(2.02) & 0.90  &  T95      &  1.26 & G91  & 0.45(0.04) &T96 \\
1226+023  & 0.158  &   5.57(1.19) & 1.40  &  T95,M97  & 44.59 & K81  &26.18(1.80) &T96 \\
\noalign{\smallskip}
\hline
\end{tabular}\\
\end{table*}

\begin{table*}
\caption{ A Sample of $\gamma$-Ray Loud AGNs with Available High Frequency Radio Data at 230~GHz.}
\begin{tabular}{lcccccccc} 
\hline\noalign{\smallskip}
 $Name$ & $Redshift$ & $N_{(\>100MeV)}(\sigma)$ & $\alpha_\gamma$ &  References & $f_{5GHz}$ & Ref & $f_{230GHz}(\sigma)$ & Reference \\
\noalign{\smallskip} \hline
1229-021  & 1.045  &   1.41(0.41) & 1.92  &  S96,T95  &  1.10 & K81  & 0.18(0.03) &T96 \\
1253-055  & 0.538  &  28.70(1.09) & 0.90  &  T95      & 16.58 & K81  &15.26(1.07) &T96 \\
1406-076  & 1.494  &  12.7 (2.34) & 1.0   &  M97,T95  &  0.5  & C97  & 0.76(0.06) &T96 \\
1510-089  & 0.361  &   4.83(1.80) & 1.3   &  T95      &  3.35 & K81  & 2.42(0.37) &T96 \\
1606+106  & 1.227  &   6.03(1.28) & 1.20  &  T95      &  1.78 & K81  & 0.64(0.30) &T96 \\
1622-297  & 0.815  &  24.56(3.18) & 1.2   &  M97      &  1.92 & K81  & 1.0        &M96 \\
1633+382  & 1.814  &  10.51(0.94) & 0.90  &  T95      &  4.08 & W85  & 1.4 (0.15) &S93 \\
1730-130  & 0.902  &  13.69(4.29) & 1.39  &  T95      &  6.90 & Gr94 & 2.61(0.19) &T96 \\
1739+522  & 1.375  &   5.38(1.11) & 1.2   &  T95      &  1.98 & K81  & 0.56(0.09) &S88 \\
1741-038  & 1.054  &   3.40       & 2.00  &  vM95     &  3.72 & K81  & 1.43(0.12) &T96 \\
1933-400  & 0.966  &   9.66(3.3)  & 1.40  &  T95      &  1.48 & K81  & 0.63(0.05) &T96 \\
2005-489$^\dagger$  & 0.071  &   1.8        &       &  vM95     &  1.50 & C97  & 0.79(0.07) &T96 \\
2052-474  & 1.489  &   3.76(2.16) & 1.40  &   T95,M97 &  2.52 & K81  & 0.66(0.05) &T96 \\
2155-304$^\dagger$  & 0.117  &   3.23(0.78) & 0.71  &  V95,M97  &  0.27 & L85  & 0.33(0.03) &T96 \\
2200+420$^\dagger$  & 0.07   &   7.81(3.83) & 1.21  &  Ca97,M97 &  4.77 & K81  & 5.4 (0.6)  &S93 \\
2230+114  & 1.037  &   4.90(1.4)  & 1.60  &  L96,T95  &  4.10 & P82  & 2.28(0.16) &T96 \\
2251+158  & 0.859  &  13.17(2.07) & 1.20  &  L96,T95  & 23.30 & W85  &10.80(0.87) &T96 \\ 
\noalign{\smallskip}
\hline
\end{tabular}\\
$^\dagger$: BL Lac object\\
References: \\ B91: Becker et al. (1991); 
Ca97: Catanese et al.(1997); C97: Comastri et al(1997); 
F94: Fichtel et al.(1994); G91: Gregory \& Condon (1991); G94: Gear et al. (1994);
Gr94: Griffith et al. (1994); K81: K$\ddot{u}$hr et al. (1981); L85: Ledden \& O'Dell (1985);
L96: Lin et al. (1996); M96: Mattox \& Wagner (1996); M97: Mukherjee et al. (1997);
P82: Perley (1982); S88: Steppe et al.(1988); 
S91: Stickel et al. (1991); S92: Steppe et al.(1992); S93: Steppe et al.(1993); S96: Sreekumar et al.(1996); 
Sh96: Shrader et al.(1996); T95: Thompson et al.(1995); T96: Tornikoski et al.(1996); 
V95: Vestrand et al.(1995); vM95: von Montigny et al.(1995); W85: Wall \& Peacock (1985)
\end{table*}

\subsection{Analysis Results}

 The observed photons are converted to flux densities at 1GeV. It is done as follows: If the the 
 photon density is expressed as $n(\nu) = n_{0} \nu^ {-(\alpha_{\gamma}+1)} $ , then the flux 
 density can be expressed as $f_{\nu} = n(\nu)h\nu \propto n_{0}\nu^{-\alpha_{\gamma}}$. $n_{0}$ 
 can be determined from the observation result  ( $N~$ photon/cm$^{2}$/s),  $N~$ photon/cm$^{2}$/s
 should be equal to  $\int_{100MeV} n(\nu) d\nu $. So, we obtained a formula to convert the 
 observed photons to the flux densities at 1 GeV, $$f_{1GeV} (pJy)=N_{(>100MeV)} \alpha_{\gamma} 10^{(2-\alpha_{\gamma})} $$
  where $N_{(>100MeV)}$ is in a unit of $10^{-7}$~photons/cm$^{2}$/s. The flux densities  are 
 k-corrected according to $f_{\nu} =f_{\nu}^{ob.}(1+z)^{\alpha-1}$, where $\alpha$ is the 
 spectral index at the  frequency $\nu$ ($f_{\nu} \propto \nu^{-\alpha}$).  The spectral 
 index is set to  0.87 and 1.25 for BL Lac objects and FSRQ (Comastri et al. 1997) for which 
 the  $\gamma$-ray spectral index is unknown, and it is chosen to be 0.0, following
 M$\ddot{u}$cke et al. (1997) for radio band. For BL Lac object 0716+714,
 a lower limit of $z=0.3 $ has been adopted, and for 0446+112, a redshift of 1.0 has 
 been used because the redshift is about 1.0 for most objects listed in the table.  
 When the linear regression  analysis is performed on  the data, the following results
 are obtained:
\begin{equation}
 {log f_{\gamma}}={(0.15 \pm 0.02)log f_{5GHz} + (1.49 \pm 0.001)}
\end{equation} with a correlation coefficient of $ r=0.16$ and a possibility of 
 the relationship having occurred by chance $p= 36\%$.

\begin{equation}
 {log f_{\gamma}}={(0.28 \pm 0.01) log f_{230GHz} + (1.55 \pm 1.8 \times 10^{-4})}
\end{equation}
 with $ r=0.347~ (p=1.7\%)$, where $f_{\gamma}$ stands for the observed maximum 
 $\gamma$-ray flux density in $pJy$, $f_{5GHz}$ and $f_{230GHz}$ stand for the
 observed radio flux density in Jy at 5~GHz and 230~GHz respectively.  
 The results are shown in figure 1 and 2.

\begin{figure}
\epsfxsize=15cm
$$
\epsfbox{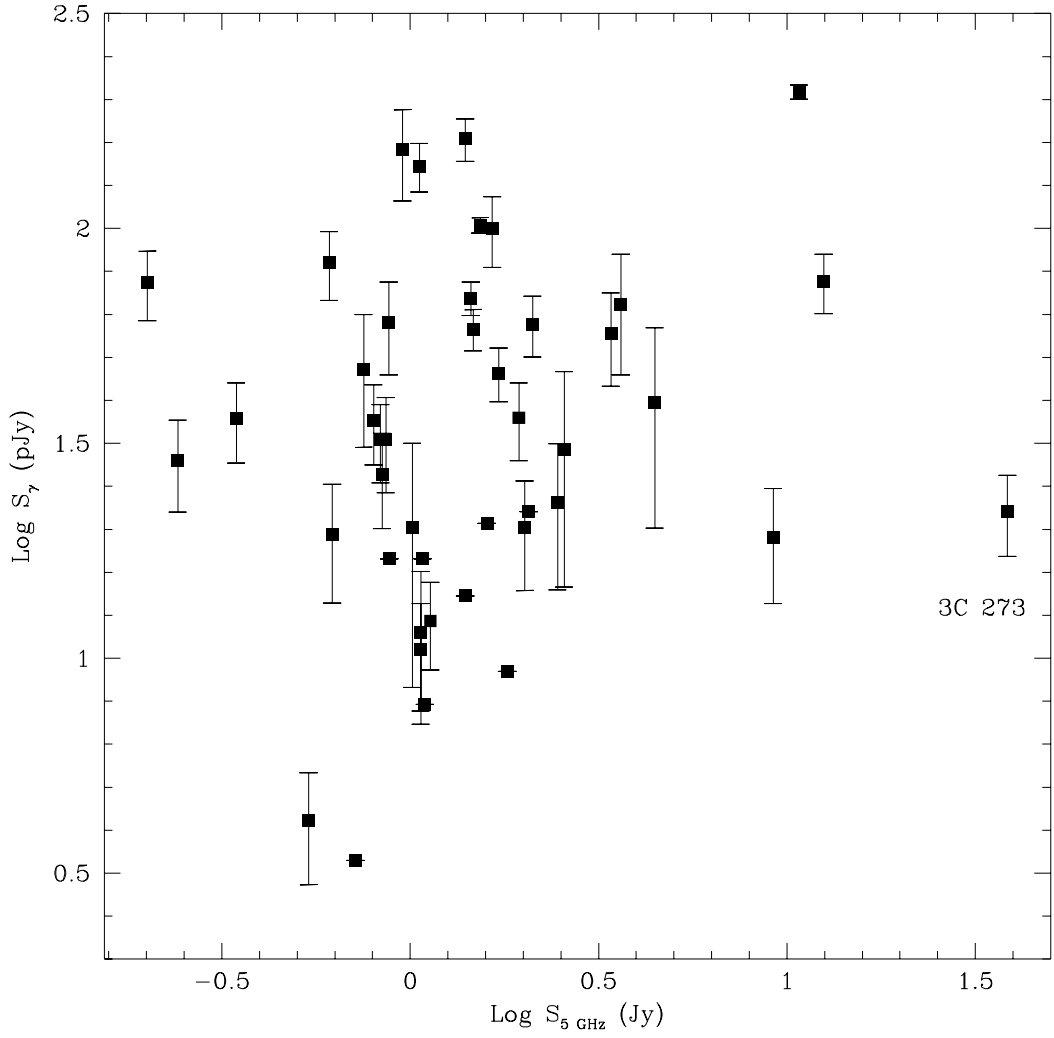}
$$

\caption{ The diagram of $\gamma$-ray flux density in pJy against the 
 radio flux density in Jy at 5~GHz}
\end{figure}

\begin{figure}
\epsfxsize=15cm
$$
 \epsfbox{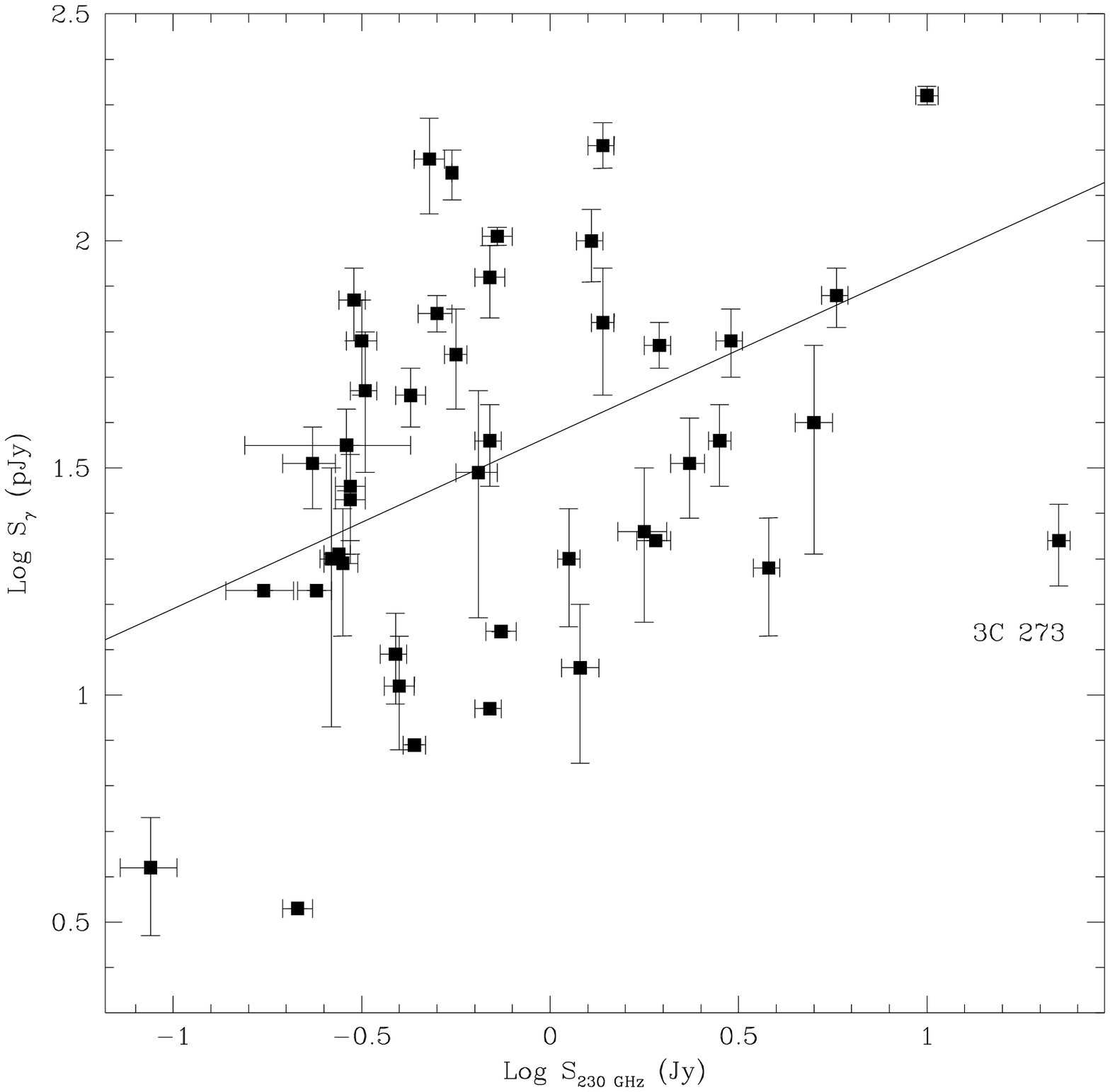}
$$

\caption{ The diagram of $\gamma$-ray flux density in pJy against the
  radio flux density in Jy at 230~GHz, the
 solid line shows the best fit  with $3C273$ excluded}
\end{figure}

\section{Discussion}

 Observations show that the $\gamma$-ray loud AGNs are clearly associated with compact, 
 flat radio spectrum sources. These objects show evidence for superluminal motion 
 (von Montigny et al. 1995).  Schachter \& Elvis (1993) reported that there is a
 correlation between the $\gamma$-ray and radio emission at 6cm (5GHz), but a negative 
 result was reported by  M$\ddot{u}$cke et al. ( 1997). We think that the problem is 
 from the facts mentioned in the introduction.  For large $\gamma$-ray flares in blazars, 
 they only occur when the sources are in a high state, and many blazars are detected 
 only in a flare state (Hartman 1996; also see McHardy 1996). So, a  $\gamma$-ray emitter 
 is more easily detected when it is in a flare state.  If the $\gamma$-ray emission is 
 from the $SSC$ model, there should be a correlation for the fluxes in the flare  
 between the radio flux and the $\gamma$-ray flux.  The $f_{\gamma}-f_{radio}$ 
 correlation places an observational constraint on the $\gamma$-ray radiation mechanism
  and can be applied to test the radiation models of  the emitting region.
  It is clear from section 2  that there is a  correlation for the maximum fluxes 
 between the $\gamma$-ray and the 230GHz bands, but the correlation between the 
 $\gamma$-ray and the radio emission at 5~GHz is weaker. 

 It is well known that both the radio radiation of blazars and the $\gamma$-ray 
 emission are strongly beamed, which means that there should be a correlation 
 between the $\gamma$-ray and the radio data in the jets, and it is hard for us 
 to get a good correlation between the $\gamma$-ray and the (5~GHz) radio band 
 since the 5~GHz radio flux is not wholly from the jet. That may be why different 
 results have been reported.

 From the figures, we can see that $3C273$ lays at bottom right, which suggests 
 that the object was not in its flare state when it was observed. If we exclude 
 this object, a  better correlation: 
 ${log f_{\gamma}}={(0.38 \pm 0.02) log f_{230GHz} + (1.57 \pm 4 \times 10^{-4})}$ 
 with $ r=0.421 (p= 5.0 \times 10^{-3})$  shows up  (see the straight line in figure 2 ), 
 which means that the  $\gamma$-ray is associated with the high frequency radio
 emission or with the radio emission in the jets and suggests that the $\gamma$-ray 
 emission is likely from the $SSC$ process in this case. From the correlation, 
 letting $\alpha_{\gamma}=1.0$, we would expect that the flare value of 3C273 is 
 about $20 \times 10^{-7}$photon/cm$^{2}$/s in the $E > $100MeV band.

 The association between the $\gamma$-ray and the radio bands has been further 
 investigated.  Recently, Valtaoja \& Ter$\ddot{a}$sranta (1995) found a correlation 
 between the initial phase of a mm-wavelength outburst and the EGRET $\gamma$-ray 
 flaring phase of high optically polarized quasars.  Our results is consistent with theirs.  

 There is a correlation for the  maximum data between the  $\gamma$-ray and the high 
 frequency radio emissions, which suggests that the high frequency radio emission 
 ( or radio emission in the jet ) is very important for $\gamma$-ray emission.
 
\begin{acknowledgements} The authors thank the anonymous referee for his/her comments 
 and the detail annotations! This work is supported by the National Scientific Foundation 
 of China(the ninth five-year important project) and the National Pandeng Project of China. 
 JHF thanks Dr. M. Tornikoski for providing their radio data.
\end{acknowledgements}

\end{document}